\documentclass[journal]{IEEEtran}
\usepackage{graphicx}
\usepackage[dvipdfmx]{}
\usepackage{wrapfig}
\usepackage{hyperref}
\usepackage{lscape}
\usepackage{multirow}
\usepackage[caption=false,font=normalsize,labelfont=sf,textfont=sf]{subfig}
\usepackage{lineno,hyperref,subfiles,multicol}
\usepackage{url}
\usepackage[top=17.8truemm,bottom=17.8truemm,left=16.5truemm,right=16.5truemm]{geometry}
\usepackage{fancyhdr}
\begin{document}

\title{A Laser-based Time Calibration System \\ for the MEG II Timing Counter }

\author{M.~Nakao, G.~Boca, P.~W.~Cattaneo, M.~De~Gerone, F.~Gatti, M.~Nishimura, W.~Ootani, M.~Rossella, \\
Y.~Uchiyama, M.~Usami, and K.~Yoshida. 
\thanks{Manuscript received December 10, 2018. This work was supported by INFN, Italy and JSPS, Japan KAKENHI Grant Number JP26000004 and JP17J04114.}
\thanks{M.~Nakao, M.~Nishimura, and M.~Usami are with the Department of Physics, The University of Tokyo, 7-3-1 Hongo, Bunkyo-ku, Tokyo 113-0033, Japan (e-mail: nakao@icepp.s.u-tokyo.ac.jp)}
\thanks{K.~Yoshida was with the Department of Physics, The University of Tokyo, 7-3-1 Hongo, Bunkyo-ku, Tokyo 113-0033, Japan (e-mail: nakao@icepp.s.u-tokyo.ac.jp)}
\thanks{G.~Boca, P.~W.~Cattaneo, and M.~Rossella are with INFN Pavia,  6-27100, Pavia, Italy, and G.~Boca is also with the Department of Physics, The University of Pavia, 6-27100, Pavia, Italy.}
\thanks{M.~De~Gerone and F.~Gatti are with INFN Genova, 33-16146 Genova, Italy, and also with the University of Genova, 33-16146 Genova, Italy.}
\thanks{W.~Ootani and Y.~Uchiyama are with ICEPP, The University of Tokyo, 7-3-1 Hongo, Bunkyo-ku, Tokyo 113-0033, Japan}
}

\maketitle
\def\headrulewidth{0pt}

\begin{abstract}
We have developed a new laser-based time calibration system for the MEG II timing counter dedicated to timing measurement of positrons. 
The detector requires precise timing alignment between $\sim\,$500 scintillation counters. 
In this study, we present the calibration system which can directly measure the time offset of each counter relative to the laser-synchronized pulse. 
We thoroughly tested all the optical components and the uncertainty of this method is estimated to be 24 ps.
In 2017, we installed the full system into the MEG II environment and performed a commissioning run. This method shows excellent stability and consistency with another method. 
The proposed system provides a precise timing alignment for SiPM-based timing detectors. It also has potential in areas such as TOF-PET.
\end{abstract}


\section{Introduction}

\IEEEPARstart{L}{epton} flavor violating decays are powerful tools to search for new physics beyond the standard model.
The MEG II experiment at Paul Scherrer Institut (PSI) in Switzerland will search for one of these decays, $\mu^+\to e^+\gamma$, with a sensitivity of $6\times10^{-14}$ improving the existing limit of almost an order of magnitude\cite{MEG2}.
The sensitivity of MEG, the first phase of the experiment, was limited by the detector performance. 
The $\mu^+$ beam intensity was reduced to satisfy the requirements of stable operation of the detectors. 
Therefore, in order to make full use of the available beam intensity of $7\times10^{7} \mu^+/\mathrm{s}$, we developed new detectors with resolutions in energy, position, and timing improved by a factor of two.
In particular, good time resolution plays an important role to reduce the dominant background that depends quadratically on the beam intensity.

\begin{figure}[!t]
\centering
\includegraphics[width=\columnwidth]{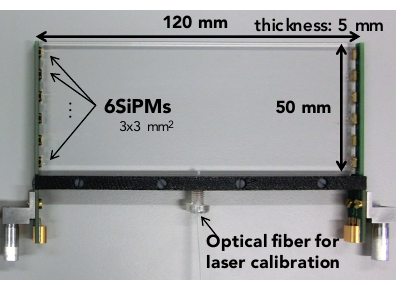}
\caption{A single scintillation counter with height 50 mm (or 40 mm depending on its position in the pTC). }
\label{counter}
\end{figure}

The pixelated Timing Counter (pTC), one of the key detectors in MEG II, is dedicated to the measurement of the positron timing. 
Its design is based on a new approach to improve the time resolution: the multiple hit scheme\cite{VCI}. 
The pTC consists of 512 scintillation counters (Fig. \ref{counter}) and is divided into two equal sectors: one placed upstream the MEG II target and the other downstream. 
Each counter, with a size of 40 (50) $\times$ 120 $\times$ 5 mm$^3$, consists of a fast plastic scintillator (SAINT-GOBAIN, BC422) read out by six series-connected silicon photomultipliers (SiPM, AdvanSiD, ASD-NUV3S-P High-Gain, 3 $\times$ 3 mm$^2$, 50 $\times$ 50 $\mu$ m $^2$, V$_{\mathrm{breakdown}}\sim\,$24\,V) at both short ends. Signal positrons hit $\sim\,$9 counters on overage and the total time resolution is expected to improve with the number of hit counters. This approach enables us to achieve a time resolution of 35 ps, two times better than in MEG.

Exploiting the multiple hit scheme requires a precise relative time calibration between counters, which is a challenging task. 
We have developed a new time calibration method using a pulsed laser and commercially available optical components. 
This method measures directly the time offsets for the timing alignment.

\section{Overview of the laser-based time calibration system}
\label{sec:laser_system}
\begin{figure*}[ht]
\centering
\subfloat[]{\includegraphics[width=\columnwidth]{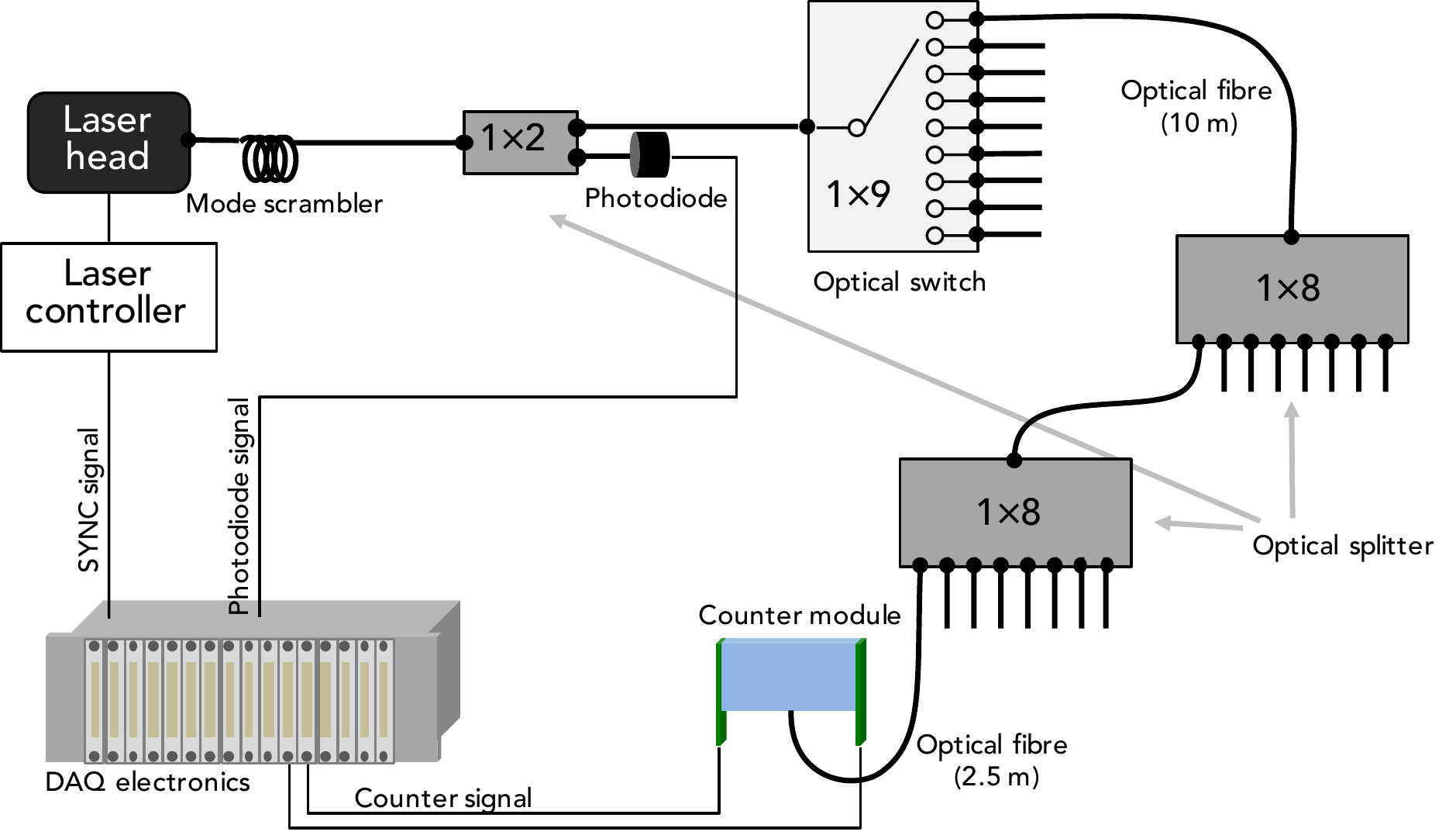}
\label{laser_system_a}}
\hfil
\subfloat[]{\includegraphics[width=\columnwidth]{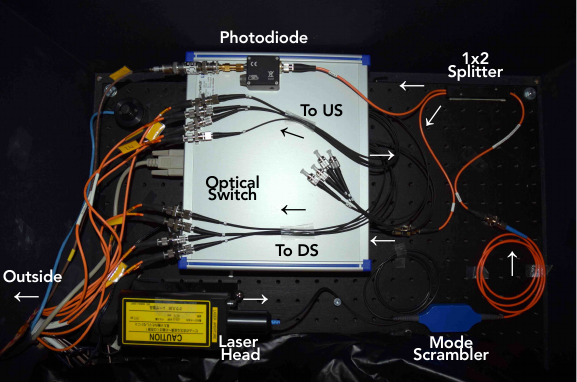}}
\label{laser_system_b}
\caption{(a)Schematic and (b)picture of the laser-based time calibration system for the MEG II timing counter. A detailed description can be found in Sec.\ref{sec:laser_system}.}
\label{laser_system}
\end{figure*}

The laser-based time calibration system is shown in Fig. \ref{laser_system}. 
The laser pulse (``Laser controller" and ``Laser head") is split into each scintillation counter simultaneously via several optical components: a mode scrambler, optical splitters, and optical fibers.
The laser device is produced by Hamamatsu with a part number of PLP-10 and has the following features: wavelength: 401\,nm, pulse duration: 60(typical)--100(max)\,ps, peak power: 200 mW.
The mode scrambler produced by OPTO SCIENCE, INC. stabilizes transmission mode in a fiber.
The optical switch produced by LEONI switches light paths.
Optical splitters divide the input light into several channels. 1 to 2 splitter is produced by OZ OPTICS and the others are produced by Lightel Technologies Inc.
We use multimode and graded index fibers with a core diameter of 50\,$\mu$m and a cladding diameter of 125\,$\mu$m.
There are produced by OZ OPTICS (for 2.5\,m) and Lightel Technologies Inc. (for 10\,m).
The selection of these items was based on a detailed study presented in \cite{Bertoni}. To get enough laser power for each counter, one-eighth of counters ($\sim\,64$) are illuminated at one data taking. The time offset of each counter is measured relative to laser-synchronized pulse. 
We have developed a new method to fix an optical fiber into the scintillator: a small hole with a diameter of 2.5\,mm and with a depth of 1\,mm is drilled into the bottom long side of the scintillator and a polycarbonate screw (at the bottom in Fig. \ref{counter}), which has a fiber-shaped hole inside, fixes the optical fiber together with the black support bar made of ABS resin.

Slow control tasks such as monitoring temperatures, controlling the laser, and switching the optical switch are performed based on the Midas Slow Control System\cite{midas} developed at PSI and TRIUMF.

\section{Research and development}
We focused on the following items in research and development of the laser calibration system.
\subsection{Fiber insertion method}
Firstly, we confirmed that the hole in the scintillator does not degrade the time resolution. 
Then we considered whether we should put optical grease inside the hole expecting better optical contact. 
However, it turned out that the time offset depends on the amount of the grease. 
The amount of grease during the assembly could not be reproduced satisfactorily. 
Therefore, we decided not to use optical grease. 
We checked the reproducibility: the fiber was inserted and the time offset was measured repeatedly. 
The reproducibility of the time offset is highly satisfying (3.8 ps in standard deviation). 
The stability was also confirmed during the commissioning phase (Sect. \ref{sec:commissioning}).

\subsection{Temperature dependence}
Time offset of all optical components depends significantly on temperature. 
We measured the temperature coefficient of the whole system to be 7.6$\pm$1.5\,ps/$^\circ$C. 
The main contribution was from the ``SYNC signal" cable (RG174/U) in Fig. \ref{laser_system_a}; we will replace it with another cable with a smaller temperature coefficient (FSJ1-50A) before the engineering run in 2018. 
In addition, the temperature in the detector area is expected to be stabilized within 1\,$^\circ$C, thus the temperature effect on the time offset will be negligible.

\subsection{Mass test}
All optical components including $\sim\,$600 fibers and $\sim\,$80 splitters, were tested before installation. Particularly the optical length in the optical splitter has a large difference among different production batches. However, it does not affect the performance because this difference of optical length up to counter are subtracted from the time offset.

\subsection{Uncertainty of the system}
Finally, the uncertainty of the laser calibration is calculated to be 24 ps by summing up all the contributions which can affect the time offset such as transit time variation inside the SiPMs, reproducibility of fiber insertion method, and statistical uncertainty in the mass test.

A detailed study of the mass test and an estimation of the uncertainty are presented in \cite{PISA}\cite{LaserPaper}.

\section{Commissioning and performance}
\label{sec:commissioning}
In 2016 one-fourth of the system was operated and we reported the result in \cite{TIPP}.
In 2017 we finished the construction and the assembly of the pTC and then we installed it into the MEG II experimental area at PSI to operate it under the MEG II $\mu^+$ beam (7$\times10^7\mu/\mathrm{s}$).

\begin{figure}[!t]
\centering
\includegraphics[width=\columnwidth]{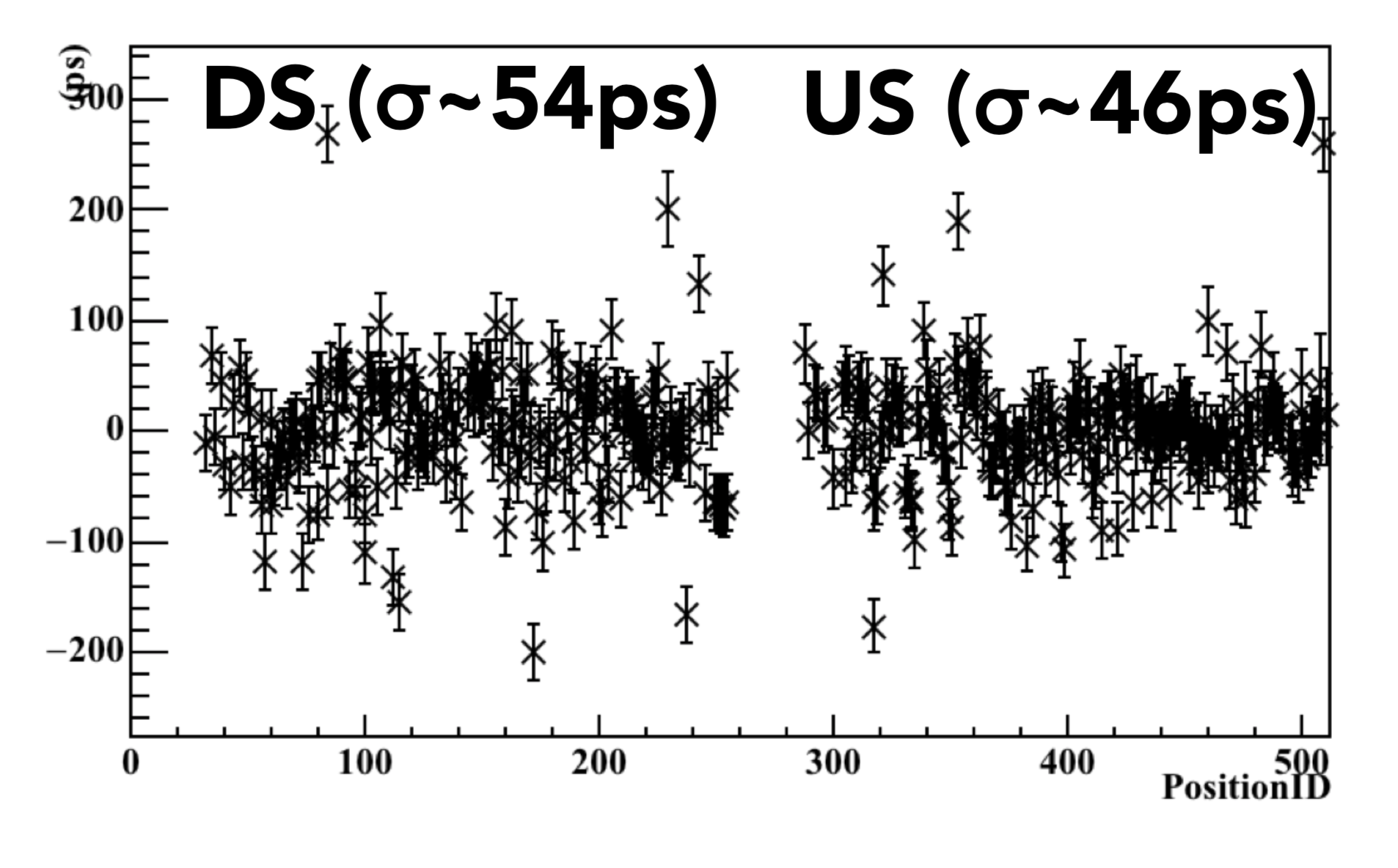}
\caption{The difference of time offsets between the laser-based method and the track-based method. Each error bar includes the systematic and statistical uncertainties of the laser-based method ($\sim\,$24\,ps). The standard deviation is 54\,ps downstream (DS) and 46\,ps upstream (US).}
\label{result}
\end{figure}

Data with positrons from Michel decay ($\mu^+\to e^+\nu\nu$) were collected for two weeks and the detector was operated for two months. 
The time offset calculated with this method was stable enough over time (2.5 ps in standard deviation for one month). 
As shown in Fig. \ref{result}, this method was compared with another calibration method, which uses positron tracks, and they are in good agreement within $\sim\,$50\,ps. 

\section{Conclusion}
We have developed the laser-based time calibration system for the MEG II pTC with an uncertainty of 24 ps. The time offset calculated with this method is stable enough for the detector operation. In addition, the consistency of the time offset was checked with another method.

Laser-based methods for TOF detectors in the past were commisioned only for plastic scintillator detectors with photomultiplier tube read-out; however, SiPM read-out enables us to design more compact and flexible timing detectors and the rapid development of SiPM technology suggests that scintillation detector with SiPM read-out will be extensively used in a wide variety of applications. 

Therefore, laser-based time calibration methods are expected to become more and more widespread. 
Furthermore, this method is precise enough to be compatible with the multiple hit scheme.
It also has potential in areas such as Time-Of-Flight positron emission tomography (TOF-PET).

The full engineering run of the MEG II experiment will start in 2019 followed by the physics run, searching for the lepton flavor violating muon decay.

\end{document}